\documentstyle[12pt]{article}
 \let\section=\subsection
 \let\subsection=\subsubsection
 \def\subsubsection#1{\subsection{#1}
   \par\note{CAUTION: subsection=subsubsection !}\par}
 
\date{}
\newlength{\lll}
\newlength{\lla}
\newcommand{\str}{\rule{0ex}{2.7ex}}  
\newcommand{\tabhline}{\\[0.3ex] \hline \str}

\newcommand{\preprint}[2]{\begin{table}[t]
            \begin{flushright} #1 \end{flushright}\vspace{#2}
            \end{table}}%
\preprint{TAUP 1920-91 \\ December 1991
}{-12mm}
\title{\vbox{\vspace{6mm}}
 Cluster Algorithm for a Solid-On-Solid Model with Constraints}
\author{\vbox{\vspace{9mm}}
   {\bf Martin Hasenbusch$^{1}$, Gideon Lana$^{2}$},\\
   {\bf Mihail Marcu$^{2,3}$ and Klaus Pinn$^{4}$}\\[9mm]
$^1\,$Fachbereich Physik, Universit\"at Kaiserslautern,\\[-1mm]
      Postfach 3049, D-6750 Kaiserslautern, Germany\\[4mm]
$^2\,$School of Physics and Astronomy,\\[-1mm]
      Raymond and Beverly Sackler Faculty of Exact Sciences,\\[-1mm]
      Tel Aviv University, 69978 Tel Aviv, Israel\\[4mm]
$^3\,$II. Institut f\"ur Theoretische Physik, Universit\"at Hamburg,
      \\[-1mm]
      Luruper Chaussee 149, D-2000 Hamburg 50, Germany\\[4mm]
$^4\,$Institut f\"ur Theoretische Physik I, Universit\"at M\"unster,\\
      [-1mm]
      Wilhelm-Klemm-Str.\ 9, D-4400 M\"unster, Germany\\[-3mm]}%
\begin{document}
\maketitle \vfill
\begin{abstract} \normalsize 
We adapt the VMR (valleys-to-mountains reflections) algorithm,
originally devised by us for simulations of SOS models, to the
BCSOS model. It is the first time that a cluster
algorithm is used for a model with constraints.
The performance of this new algorithm is studied in detail in both
phases of the model, including a finite size
scaling analysis of the autocorrelations. \\[5mm] \noindent
{\bf PACS} numbers: 68.35.-p, 68.35.Rh, 64.60.Fr, 64.60.Ht, 05.50.+q
\end{abstract} \vfill
\thispagestyle{empty}
\newpage
\section*{Introduction}

Cluster algorithms are one of the most promising new ideas to overcome
the problem of critical slowing down (CSD) in computer simulations of
statistical physics models \cite{SW,gencluster}. Each time they were
sucessful, the dramatic improvement in performance came by integrating
into the algorithm the relevant physical properties of the model under
study. In the case of solid-on-solid (SOS) models \cite{SOS}, we devised
a new cluster algorithm \cite{ouralg} that is based on reflecting
``valleys'' and ``mountains'' of the fluctuating SOS surface. We shall
call this algorithm the VMR algorithm
(valleys-to-mountains-reflections).

Up to now, no cluster algorithms were devised for {\em models with
constraints} on the configurations. In this letter we present a first
example where this is achieved: the adaptation of the VMR algorithm
to the BCSOS model.

The BCSOS model is defined as follows \cite{vanBeijeren,SOS}.
On each site $i$ of a square
lattice there is an integer-valued variable $h_i$, which can be thought
of as the height of a surface above the two-dimensional lattice. Nearest
neighbor variables are constrained to a height difference of $\pm 1$.
Next to nearest neighbors (i.e.\ diagonal neighbors) can thus have a
height difference of $\pm 2$ or $0$, and the $\pm 2$ is assigned the
smaller Boltzmann weight. The partition function is then
\begin{equation}\label{e1}
Z = {\sum_{h}}^{\prime} \exp \left( - \mbox{\small $\frac{1}{4}$}
       K \sum_{[i,j]} (h_i - h_j)^2 \right)\; ,
\end{equation}
where $[i,j]$ are pairs of diagonal neighbors ($\sum^{\prime}$ is the
constrained sum). The model undergoes a
roughening transition at $K_r=\ln 2$. At large $K$ the SOS surface is
smooth, at small $K$ it is rough. The BCSOS model is equivalent to the
F-model \cite{SOS,Lieb,Baxter}, which is a particular case of the
6-vertex model (the coupling $K$ in eq.~(\ref{e1}) is chosen such as to
conform with standard 6-vertex notations).

The idea for the VMR algorithm came from the picture of the SOS
configurations as landscapes with mountains and valleys. We first choose
a horizontal {\em reflection plane}. The clusters are, roughly, the
connected regions above the plane (mountains) and below it (valleys).
Large scale changes of a configuration are performed by ``flipping''
valleys or mountains, i.e.\ by reflecting them through the plane
independently and with an appropriate probability. Then a new reflection
plane is chosen, the reflection procedure repeated, and so on.

In one way or another the reflection idea had turned up in other cluster
algorithms \cite{Brower,Wolff,gencluster}. In the case of the VMR
algorithm for SOS models however, the crucial ingredient in eliminating
CSD altogether turned out to be not only the reflection, but the precise
procedure for choosing the reflection plane \cite{ouralg}. It is at
this stage that the particular features of the SOS models were
incorporated into the algorithm.

We shall see that the VMR algorithm for the BCSOS model eliminates CSD
only half-way. It is nevertheless of considerable algorithmic
significance to achieve even that much for a model with constraints.
Besides, the improvement in performance was enough for the purpose of a
high precision determination of the roughening transition in several
SOS models \cite{KTTc}. In that study the renormalization group flows of
the SOS models were matched to that of the critical trajectory of the
BCSOS model.

\section*{The VMR algorithm for the BCSOS model}

We shall describe a ``single cluster flip'' algorithm
\cite{Wolff,gencluster}. A step consists in growing a cluster around a
seed, then flipping it. The seed is a randomly chosen site $i_0$.

For a given seed $i_0$, the first thing is to choose an {\em integer
reflection plane} $M$. We tried two choices of $M$. One is to set $M = h_{i_0}
\pm 1$, the $+1$ or $-1$ being chosen randomly with probability one half.  The
other is to choose another random site $j_0$, and set $M=h_{j_0}$. Both choices
gave the same dynamical exponent $z$ (see definition below). The smallest
autocorrelation times were obtained with the second procedure, in the case that
$j_0$ is taken on the sublattice that does not contain $i_0$ (we have an even
and an odd sublattice; if on the even sublattice all heights are even integers,
then on the odd sublattice they are odd integers, and vice versa).

Starting from the seed, which is the first site in the cluster, we {\em
freeze} and {\em delete} links with probabilities to be given below.
We have to consider both nn links
(nearest neighbor) and nnn links (next to nearest neighbor). A new site
is taken into the cluster if it is connected by a frozen link to a site
already in the cluster. We repeat the freeze/delete procedure until all
links pointing from sites in the cluster either end in the cluster or
are deleted. After the cluster is finished, we reflect it (flip it)
through the reflection plane $M$:
\begin{equation}\label{e2}
 h_i \rightarrow 2 M - h_i
\end{equation}
for all sites $i$ in the cluster.

The freeze/delete probabilities are defined as follows (for each link
the sum of the freeze and delete probability is one). For the nn link
$<i,j>$ we have two situations. If either $h_i=M$ or $h_j=M$, the link
is always deleted. Otherwise (i.e.\ if both sites are on the same side
of the reflection plane) the link is always frozen. Thus it is ensured
that the constraint on the nearest neighbors is not violated by the
reflection.

As a consequence of the freeze/delete procedure for nn links,
the nnn link $[i,j]$ is automatically frozen in all but three
situations (remember that $|h_i-h_j|$ is zero or two). First,
if $h_i$ and $h_j$ are on different sides of the reflection plane,
the nnn link $[i,j]$ is always deleted.
Second, if either $h_i=M$ or $h_j=M$, $[i,j]$ is also deleted with
probability one.
Third, if $h_i = h_j = M \pm 1$, the nnn link $[i,j]$ is deleted with
probability $\exp(-K)$.
In eq.~(\ref{e1}) we wrote the nnn interaction in the form of the discrete
Gaussian model. It can be easily seen that the delete probabilities
used for the nnn links are indeed the same as those defined in
\cite{ouralg} for the discrete Gaussian model without constraints.

Notice that for the case of the BCSOS model, the clusters are precisely
valleys or mountains cut by the reflection plane from the SOS surface.
In the case of the discrete Gaussian model \cite{ouralg}, this was only
approximately so.

\section*{Autocorrelations, sublattice effects}

We shall call ``autocorrelation time'' $\tau$ the quantity often referred
to in the literature \cite{gencluster} as the ``exponential autocorrelation
time''. At large ``Monte Carlo times'' $t$ the autocorrelation function of
some physical quantity decays exponentially as $\exp(-t/\tau)$, which is
related to the slowest mode in the Markov process used for the simulation.

In practice, for some physical quantities the autocorrelations
immediately enter the $\exp(-t/\tau)$ regime, while for other quantities
this only happens after the autocorrelations have decayed much faster
for a while. Of course, the theoretical expectation is that at very
large $t$ the autocorrelations decay with the same $\tau$ for all
quantities (except if prevented by some conservation law).

For the BCSOS model we found that quantities defined on one
sublattice tended to behave as $\exp(-t/\tau)$ already for $t$
of the order of a few to a few tens of cluster flips. On the other hand,
averages over the two sublattices tended to enter this ``clean exponential''
regime only after the autocorrelation functions had decayed over several
orders of magnitude.

As an example we discuss here the energy density, which
is, up to a constant factor, $<V^{-1}\sum_{[i,j]}(h_i-h_j)^2>$, with $V$ the
number of points on one sublattice, and the sum over nnn pairs $[i,j]$ in that
sublattice. The values of the energy density on the two sublattices are
so strongly anticorrelated that the statistical error of the average over the
two sublattices is a whole order of magnitude less than the error of
the one-sublattice quantity.

These strong anticorrelations can be understood by regarding the four points
at the corners of an elementary plaquette of the lattice.
In the case that the height difference of the nnn variables at one pair of
diagonally opposite sites is two,
the constraint on the nn pairs implies that for the other pair of diagonally
opposite sites the height difference is necessarily zero.
The anticorrelations stem from the fact that the two nnn pairs discussed here
lie on different sublattices.

In what follows, we shall present the results for the autocorrelation time of
the one-sublattice energy. At each value of $K$ and for each lattice size
quoted, the statistics was between 250000 and 800000 clusters (vectorization of
the cluster algorithm \cite{HGE} was of great help). We used periodic
boundary conditions in the directions of the coordinate axes of the sublattices
(these axes are rotated by $45^{\circ}$ with respect to the coordinate axes of
the original lattice). Each sublattice has a size of $L^2$, so our lattice has
volume $2L^2$. For our study we took $L$ to be a power of 2, and
 $8 \leq L \leq 256$
(in the rough phase the largest $L$ was 128).

As opposed to the one-sublattice energy,
for the energy averaged over both sublattices
the $\exp(-t/\tau)$ regime is not reached before
the autocorrelations become zero within errorbars, if
the statistics is a few hundred thousand of clusters.
In some runs with ten times as many clusters we checked that the energy
averaged over the two sublattices does indeed have the same $\tau$.

The values for $\tau$ we shall quote are in units of sweeps, i.e.\ we compute
$\tau$ in units of clusters first, then multiply it by the average cluster size
and divide it by $2L^2$. One unit then amounts to a work proportional to the
volume. As opposed to other cluster algorithms as e.g.\ the single
cluster algorithm for the Ising model \cite{Wolff,gencluster}
 the average cluster size is quite
stable. If we choose the reflection plane to be the height at a random site on
the sublattice not containing the seed, the average cluster size is around
0.35 of the volume.

\section*{Autocorrelation study in the rough phase and at $K_r$}

Since the SOS surface thickness increases only logarithmically with the
lattice size, the SOS surface has in reality shallow valleys and
small mountains. As the temperature increases ($K$ decreases) the
valleys become deeper and the mountains higher. This leads us to the
expectation that the higher the temperature, the better our algorithm
should perform.

We studied the autocorrelations at the roughening transition $K_r = \ln 2$ and
at the point $K=K_r/2$ which is deep in the rough phase. Since in the rough
phase the correlation length is infinite, the dynamical critical exponent $z$
for CSD is defined by the relation \cite{gencluster}
\begin{equation}\label{e3}
 \tau \propto L^z \; .
\end{equation}
The results are shown in figs.\ 1 and 2.
Fig.~1  suggests that at $K_r$ eq.~(\ref{e3}) is well
fulfilled even for small $L$. The best fit of our data gives
\begin{equation}\label{e4}
 \tau(K_r) = 1.20 \pm 0.02 \; .
\end{equation}
At $K_r/2$, fig.~2 suggests that the regime of eq.~(\ref{e3})
was not yet reached, and that by increasing $L$ further the fitted value of
$\tau$ would decrease. We conclude that
\begin{equation}\label{e5}
 \tau(K_r/2) < 0.79 \pm  0.09 \; ,
\end{equation}
the number on the r.h.s.\ being the fitted value for $32 \leq L \leq 128$.
Thus the algorithm does indeed perform better at higher temperatures.

\begin{figure}
    \vspace{8cm}
    \caption[fig1]{Log-log plot of $\tau$ vs.\ $L$ at $K=K_r$.
       The solid line is the fit with
       eq.~(\ref{e3}) for $8\leq L\leq 128$.}
    \label{fig1}
\end{figure}

\begin{figure}
    \vspace{8cm}
    \caption[fig2]{Log-log plot of $\tau$ vs.\ $L$ at $K=K_r/2$.
       The solid line is the fit with
       eq.~(\ref{e3}) for $32\leq L\leq 128$.}
    \label{fig2}
\end{figure}

In \cite{ouralg} we analyzed the VMR algorithm for the Discrete Gaussian
model. The BCSOS algorithm presented here performs similarly to the
H-algorithm discussed there. Because of the constraints, we did not find
a way to include the analogue of the
I-algorithm that further improved the performance
in the case of the Discrete Gaussian model.

\section*{Finite size scaling of the autocorrelation time in the smooth phase}

In \cite{SokalXY} it is conjectured that the autocorrelation time obeys
the usual finite size scaling law
\begin{equation} \label{e6}
\tau = L^z \, f(L/\xi) \; ,
\end{equation}
where $f$ is some smooth function.
We therefore took data for fixed ratios $L/\xi$. Fortunately, there is an
exact formula for the correlation length of the BCSOS model
(ref.\ \cite{Baxter}, eq.~8.11.24).
We made runs for $L/\xi = $ 0.25, 0.5, 1, 2, 4 and 8,
with $L$ chosen to be a power of two, $8 \leq L \leq 256$.

For each value of $L/\xi$, we fitted the values of $\tau$, obtained from the
autocorrelations of the one-sublattice energy, with eq.~(\ref{e3}). In each
case the fits were good, but, as opposed to eq.~(\ref{e6}), the value of $z$
was not constant. In table~1 we give the fitted values of $z$ as a function
of $L/\xi$. We see that $z$ slowly increases as $L/\xi$ decreases, and this
increase seems consistent with the value $z=1.20(2)$ at the transition point
($L/\xi=0$).

\begin{table}
 \centering
 \caption[dummy]{\label{tab1} The exponent $z$ for fixed $L/\xi$ in the
   smooth phase.}
 \vspace{2ex}
\begin{tabular}{|c||c|c|c|c|c|c|c|}
\hline\str
 $L/\xi$ &   0  &  0.25  &  0.5  &  1  &  2  &  4  &  8
\tabhline
  $z$ & 1.20(2) & 1.08(4) & 1.02(5) & 0.90(4) & 0.91(4) & 0.86(5) & 0.77(9)
\\[.3ex] \hline
\end{tabular} \end{table}

This finding is compatible with a scaling law $\tau \propto \xi^z$ in the limit
$L\rightarrow\infty$, but with an exponent $z$ smaller than 1.20(2).
In all cases investigated up to now the two exponents were equal, so such a
situation would be very interesting. We stress, however, that we do not have
data close enough to the thermodynamic limit in order to claim that the law
$\tau\propto\xi^z$ was well checked.

In some cases logarithmic corrections to the finite size scaling variable
$L/\xi$ turn out to be important \cite{Brezin}.
We tried for the same data for $\tau$ in the smooth phase the ansatz
\begin{equation}\label{e7}
\tau = L^z \, f\!\left(\frac{L}{\xi(\ln{\xi})^a}\right) \; ,
\end{equation}
where we took $z = 1.20$. We looked for a value of $a$ for which the
values of $\tau/L^z$ plotted against $L/(\xi(\ln{\xi})^a)$
collapse onto one curve. In fig.~3 we show the result for
$a = -0.7$. The range of $a$ for which reasonable
collapse occured was $ -0.9 \leq a \leq -0.6$.
Thus, within our precision, this
modified finite size scaling is consistent with all data and with the
value $z=1.20$. We caution however that a collapse of the data of
roughly the same quality can be
found for $ 1.05 \leq z \leq 1.25$.

We do not claim that eq.~(\ref{e7}) is the ``correct'' finite size scaling
law, but it is certainly closer to the truth than eq.~(\ref{e6}).

\begin{figure}
    \vspace{10cm}
    \caption[fig3]{Finite size scaling plot with $z=1.20$ and $a=-0.7$.
                   Notice the logarithmic axes.
                   Different marker symbols are used for different
                   values of $L/\xi$.}
    \label{fig3}
\end{figure}

\section*{Conclusions and outlook}

We have presented for the first time a successful cluster algorithm for
a model with constraints. We carefully measured and analyzed the
performance of our algorithm. We also performed a finite size scaling
analysis for the autocorrelation time.

We would like to point out that a completely different algorithm for
vertex models was also developed recently \cite{newvertex}.
Another research direction we pursue is the generalization of the VMR
algorithm to more complicated constraints, which may be of relevance
for comparing to interface experiments.

While further algorithmic studies are interesting by themselves, we
point out that the algorithm described in this letter was used in the
(to date) most precise determination of the roughening
(Kosterlitz-Thouless) transition for a variety of SOS models
(including the dual of the XY model) \cite{KTTc}.

Finally, we point out that our cluster algorithm can be used, in
slightly modified form, for accelerating
quantum Monte Carlo simulations for XXZ chains \cite{Suzuki}.

\section*{Acknowledgements}

This work was supported in part by the German-Israeli
Foundation for Research and Development (GIF), by the Basic Research
Foundation of The Israel Academy of Sciences and Humanities, and by the
Deutsche Forschungsgemeinschaft. We would like to express our gratitude
to the HLRZ in J\"ulich, where most of our computer runs were performed.

\end{document}